\documentstyle[prb,aps,twoside,floats,epsf]{revtex}

\begin{document}

\twocolumn[\hsize\textwidth\columnwidth\hsize\csname
@twocolumnfalse\endcsname

\draft 
\preprint{Submitted to Phys. Rev. B}

\title{Single-Electron Charging and Coulomb Interaction in InAs Self-Assembled Quantum Dot Arrays}

\author{G. Medeiros-Ribeiro}

\address{Materials Department, University of California, Santa Barbara
CA 93106}

\author{F. G. Pikus}

\address{Department  of  Physics,  University  of  California,   Santa
Barbara CA 93106}

\author{P. M. Petroff}

\address{Materials Department, University of California, Santa Barbara
CA 93106}

\author{A. L. Efros}

\address{Department of Physics, University of Utah, Salt Lake City, UT
84112}

\date{\today}
\maketitle

\begin{abstract}
Sequential single-electron charging is observed in InAs Self-Assembled Quantum
Dots using capacitance
spectroscopy. In this system, the Coulomb energy is
smaller than the inter-level
energy spacings due to the quantum confinement and both effects can
be separately identified.
A theoretical model is
proposed for this system and the capacitance experiments were
devised in order to
experimentally observe the effects of Coulomb interaction
between electrons on the dots.
The effects
of inter- and intra-dot Coulomb interaction have been observed in the capacitance
spectra.
A good
agreement between the
proposed model and experiment is achieved.
\end{abstract}

\vskip 2pc ] 

\narrowtext

\section{Introduction}

Quantum effects and Coulomb  effects in zero dimensional  systems have
been extensively studied in the past  years due mainly to the new  and
interesting phenomena these structures exhibit\cite{1}. The degree  of
importance of each  of these effects  will depend on  the size of  the
objects that form the system,  since  Coulomb  effects  should  diminish  linearly  with the
inverse  of  the  dimensions  of  the  system  whereas quantum effects
diminish quadratically. Artificially  produced systems, such  as those
defined  by  electron  beam  lithography\cite{2}  and other patterning
techniques as well  as in structures  with split gate  geometry\cite{3},
are  usually  constrained  by  technological limitations. The ultimate
size  one  obtains  with  such  techniques  is still too large to make
quantum energies comparable to  thermal energies beyond the  mK range,
and as a rule Coulomb energies prevail in these systems. Nevertheless,
systems where the growth  kinetics naturally produce small  structures
offer the possibility of decreasing even further the dimensions. As an
example, islanding in  highly strained heteroepitaxial  coherent III-V
systems\cite{4}  produces  structures  with  dimensions  were  quantum
effects  are  routinely  observable\cite{5,6},  and  furthermore,   at
elevated     temperatures\cite{7,8}.     In     this     work,     the
InAs self-assembled  quantum dots grown on GaAs\cite{7,8}  
are  investigated  with  capacitance   spectroscopy.
Previous works  on this  system using  the capacitance  technique were
focused  on  the  magneto-optical  properties\cite{9}  and  electronic
structure\cite{10} of the  dots. The focus  on the present  work is on
the electrostatic interactions  in this system,  where we analyze  the
effect of having a finite number of electrons in a single dot and  the
adjoining  consequences  on  the  capacitance  spectra of having their
correspondent image  charges in  a nearby  electrode and  electrons in
neighboring dots.  
It is demonstrated that the static characteristics
show the independent loading of single electrons in arrays of  about
$10^9$
dots.
We successfully
predicted  and  simulated  the  effects  of  the 
Coulomb   interaction   between electrons in the dots with the gate and
electrons in other dots.
The structure of this paper is as follows:
in the experimental section
the growth  and the  capacitance experiment  will be  described, along
with the identification of the components inherent to each capacitance
spectrum. The theoretical model  section describes the model  utilized
to predict and simulate  the capacitance spectra and  the electrostatic
effects, and the results and  analysis section presents the  measured data
and  the  simulation  results.

\section{Experiment}

\begin{figure}[t]
\epsfxsize=3in
\epsffile{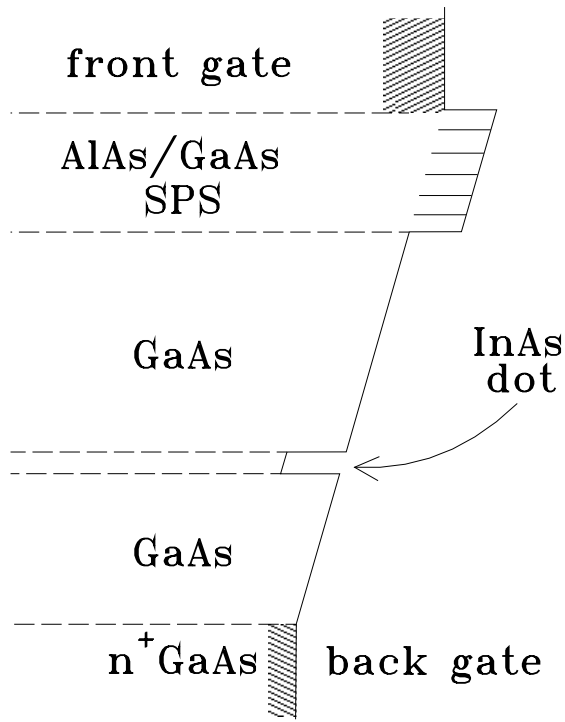}
\caption{Sketch of the band diagram of the quantum dot array sample in
the direction perpendicular to the dot and gate planes.
\label{FigBand}}
\end{figure}

The samples studied  in this work  were grown on  a Varian GEN-II  MBE
system  on  semi-insulating  [100]  oriented  2"  GaAs substrates.
The conduction band diagram of a sample is shown schematically in
Fig.~\ref{FigBand}. The
growth procedure was the same as used in other  works\cite{5,9,10,12},
consisting on the oxide  desorption under $10^{-5}$ Torr  ${\rm As}_4$
flux and subsequent growth of a 1 micron undoped GaAs buffer layer  at
1 ML/s at 620 $^\circ$C. We also used  40 periods of a $2 {\ \rm nm}  \times 2
{\ \rm nm}$ AlAs/GaAs short period superlattice (SPS) to trap  defects
and smooth the  surface prior to  the deposition of  a 20-100 nm  GaAs
undoped  layer,  20-80  nm  thick  $2  \times 10^{18} {\ \rm cm}^{-3}$
n-doped  back  contact  and  a  undoped  GaAs  tunneling  barrier with
thickness $d$. The growth temperature is now reduce to 530  $^\circ$C\cite{12}
and InAs is deposited, with  the substrate fixed with the  $1\bar{1}0$
direction aligned  with the  axis of  the In  cell. By  doing that, we
could produce a variation on the deposited thickness of InAs of up  to
10\%,  assuming  an  ideal  point  source  for  the effusion cell. This
procedure allowed us  to obtain a  
the dot density
 ranging
from 0 in one side of the  wafer up to $2 \cdot 10^{10} {\ \rm  cm^{-2}}$, as
estimated by  TEM and  capacitance measurements.  The growth  was then
resumed by depositing GaAs at  530 $^\circ$C with the substrate  now rotating.
After 5  nm of  GaAs thickness,  the temperature  was increased  to 600 $^\circ$C
remaining at this value for the remaining part of the growth. A 25  nm
thick GaAs  was the  deposited, and  another AlAs/GaAs  $3 {\  \rm nm}
\times 1 {\ \rm nm}$ SPS was grown before capping the
sample with a 5 nm thick GaAs layer.

When analyzing  the experimental  data, one  should take  into account
that the potential in  the dot plane $V$  is not equal to  the
voltage  on the gate $V_g$.  If the dots are located at  a distance
$d$ from the  back gate and  the distance between  the front and  back
gates is $t$ than

\begin{equation}
\frac{V_g}{V} = L \equiv \frac{t}{d}.
\label{lever}
\end{equation}

\noindent We  call the  ratio $L$  a {\em  lever-arm coefficient}, it
varies from 6 to 7 in different samples.
We kept values of $L$ similar for all samples with different $d$ by
varying the thickness of the top AlAs/GaAs SPS.

Samples with tunneling barriers thinner than 20 nm were not used since
the strain field  produced by the  dots extends into  the substrate for
approximately this  amount, as  qualitatively inferred  by changes  in
contrast  in  cross-section  TEM  micrographs.  The  thickness  of the
back-contact had also to be  increased so that the strain  field would
not  deplete  the  regions  beneath  the  dots  and therefore producing
effectively different d values.

Metallic disks were evaporated after contacting the back contact layer
using standard procedures for making  ohmic contacts in n doped  GaAs.
The photolithographically defined disks consisted of a thin layer of Cr  (5
- 10 nm) followed by a 200  nm thick Au layer, with diameter from  150
$\mu$m  to  350  $\mu$m.  The diodes were bonded and  their capacitance was
measured  at  4.2 K in a liquid  He  immersion  cryostat.  To  obtain   the
capacitance voltage characteristics of the devices, we used a  lock-in
amplifier  at  frequencies  varying  from  40  Hz  to  30  kHz. The ac
amplitude was kept smaller than 1 meV, after conversion into energy by
the lever arm coefficient. Depending on the thickness of the  tunneling
barrier  $d$  we  used  higher  or  lower  frequencies,  choosing  the
appropriate value to have the loading and unloading of the dots within
one period of the ac frequency, i.e., $f < 2\pi R C$, where $R$ is the
tunneling resistance and $C$ is the capacitance determined by  the
area of each dot and the distance $d$ from the back contact.

\begin{figure}[t]
\epsfxsize=3in
\epsffile{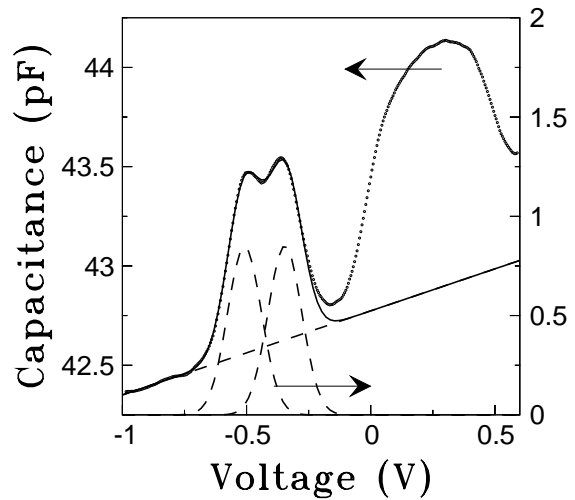}
\caption{Dot capacitance {\it vs\/} gate voltage
for the
samples with the distance $d = 450  {\rm \AA}$ between dot and gate planes.
Circles show the experimental data, solid line gives the fit by the
sum of the linear background and two Gaussian peaks (shown by dashed
lines).
\label{FigExp}}
\end{figure}

In Fig.~\ref{FigExp} we show the measured capacitance-voltage
characteristics of the sample {\sf  A} with $d = 450 {\rm \AA}$.
One can see the two peaks, corresponding to the loading of the ground
and first excited states of the dots. The first peak is split into
two; this splitting results from the Coulomb interaction of the
electron in the same dot. In this paper we analyze only this
split
peak, related to the ground state in the dots.

The following problem complicates a direct comparison of the
experimental results with theoretical capacitances:
the experimentally measured capacitance $C_{\rm exp}$ contains a large
background capacitance $C_{\rm b}$, to which the capacitance of the
dots $C_{\rm dot}$ is a small correction:

$$
C_{\rm exp} = C_{\rm b} + C_{\rm dot}.
$$

\noindent  We  resolve  this  problem  by assuming a linear background
capacitance $C_{\rm  b}$ and  extrapolating it  using the  low-voltage
part  of  the   CV  curve.  For   the   sample  {\sf  A}
we have also  measured the capacitance at  high
frequency of the ac  voltage. In this case  the electrons do not  have
the time to  tunnel to  and from  the dots  during one  period of  the ac
voltage,  and  the   dots  essentially  do   not  contribute  to   the
capacitance, whereas the background capacitance has no intrinsic  slow
processes,  and  does  not  depend  on  the  frequency in a wide
frequency range.  Thus measured
background  capacitance  shows  a  linear  voltage dependence for the voltages
where at low frequencies the dot capacitance is observed. We have also
used a  background capacitance  in the  form of  the capacitance  of a
depletion layer, which is formed in the doped back contact:

$$
C_{\rm b}^\prime = \frac{S V_d}{(V - V_0) d} \left[
  \sqrt{1 + \frac{V - V_0}{2 \pi V_d}} - 1\right],
$$

\noindent where $V_d = e^2 N_d  / \kappa$, $N_d$ is the density of the
positive charge
in the depletion layer, $\kappa$ is the dielectric constant,
$V_0$ is the voltage of flat bands, and $S$ is
the area of  the sample.
In the range of $V$ where the 
capacitance peak is observed (and for smaller voltages), 
this function can be considered as a linear one.
All  three ways of  extracting the background
capacitance lead to practically the same results.

After the background is subtracted, the capacitance correction due to
quantum dots $C_{\rm dot}$ can be very well described by a sum of two
Gaussian peaks with centers at voltages $V_1$ and $V_2$ and widths
$\sigma_1$ and $\sigma_2$, respectively:

\begin{equation}
C_{\rm dot} = \frac{Q_0}{\sqrt{2 \pi}} \left[
\frac{1}{\sigma_1} {\rm e}^{-\frac{\displaystyle (V - V_1)^2}
{\displaystyle 2 \sigma_1^2}} +
\frac{1}{\sigma_2} {\rm e}^{-\frac{\displaystyle (V - V_2)^2}
{\displaystyle 2 \sigma_2^2}} \right].
\label{gausfit}
\end{equation}

\noindent This  fit reflects  the fact  that the  areas of  both peaks
$Q_0$ should be equal,  since they give the  total charge of all  dots
with one electron per dot.  The fitting was done by  weighted explicit
orthogonal  distance  regression  using  the  software package ODRPACK
\cite{ODR}. The resulting decomposition of the experimental curve into
linear background and two Gaussian peaks is shown in
Fig.~\ref{FigExp}.

\section{Theoretical Model}

We now present a theoretical description of the quantum dot array in the
framework of the following Hamiltonian:

\begin{eqnarray}
H = &&\sum_i n_i \varphi_i +
    \frac{e^2}{\kappa}\sum_i \frac{n_i (n_i - 1)}{2 C_i} -
    \frac{e^2}{\kappa}\sum_i \frac{n_i^2}{4 d} +
\nonumber \\
    &&\frac{e}{\kappa}\sum_{i < j} n_i n_j V(r_{ij}) -
    e V \sum_i n_i.
\label{H}
\end{eqnarray}

\noindent Here the indices $i$, $j$ number the quantum dots. $n_i$  is
the occupation number of the dot  $i$; in this paper we only  consider
the ground state on the  dot, so $n_i$ can be  0, 1, and 2. The  first
term represents the non-Coulomb  disorder due to fluctuations  of size
quantization  energies  $\varphi_i$  in  the  dots.  We  assume here a
Gaussian distribution for $\varphi$:

\begin{equation}
\omega(\varphi) = \frac{1}{\sqrt{2 \pi}\sigma_\varphi}
                  {\rm e}^{-\frac{\displaystyle\varphi^2}{\displaystyle2 \sigma_\varphi^2}}.
\label{distphi}
\end{equation}

\noindent  The  width  $\sigma_\varphi$  describes  the  magnitude  of
disorder. It is a fitting parameter  in our model. The second term  in
the Hamiltonian Eq.~(\ref{H}) is the charging energy of the dot  (this
term is responsible  for Coulomb blockade-like  effects). $C_i$ is  the
capacitance of the dot, defined so that $$ W = \frac{e^2}{\kappa  C}$$
is the energy of Coulomb interaction  of two electrons in the dot\cite{WhyOne}.  
The product
$n_i(n_i-1)$ takes into account that electron does not interact with
itself. 
In
our model, we take $C_i = C$ to  be the same for all dots, and use  it
as  another  fitting  parameter\cite{whyC}.  The  third  term  in  the
Hamiltonian is the attraction energy between the charge of the dot and
its image in  the metallic electrode  at a distance  $d$ from the  dot
plane. We have considered only the attraction to the closest  metallic
electrode, which is the  back gate in our  samples. The front gate  is
much further away from the dots, and has no significant effect on  the
electrostatics near the dots. The  last term introduces the effect  of
the   gate   bias   $V$   {\bf   (}scaled   by   the   lever-arm  ratio
Eq.~(\ref{lever}){\bf)}.

The fourth term  of the Hamiltonian  Eq.~(\ref{H}) takes into  account
the  interaction  between  the  charges  on  {\em different} dots. The
interaction potential  $V(r)$ includes  the screening  of the  Coulomb
interaction by the gate and has the form:

\begin{equation}
V(r) = \frac{e}{\kappa}\left[\frac{1}{r} - \frac{1}{\sqrt{r^2 + 4
d^2}}\right].
\label{Vij}
\end{equation}

\noindent Here  we again  neglect the  effects of  the second,  remote
gate.  We  should  also  mention  that  we  have  assumed  the
point charge
 Coulomb
interaction of the dots with the  gate and with each other  
 The latter assumption is well justified when the dot
size is much smaller than the distance between the dots, which is  the
case in all  of our samples.  For the dot-gate  interaction, a trivial
calculation shows that the interaction of an electron wave function in
a disk of size $a$ with the gate at a distance $d$ differs  negligibly
from that of a point charge when $d \ge a$. This condition also  holds
for all of samples for which we present the results here.

The  thermodynamic  properties  of  the  many-electron system with the
Hamiltonian Eq.~(\ref{H}) in the regime when the Coulomb  interactions
are  essential  are  rather  complex,  and  a comprehensive solution is
possible only by means of numerical modeling. The modeling consists of
the following  steps: \\  1) Generation  of dot positions and energies.  The given number of
dots $N$ are placed inside a  square of the size $L$. The  dot density
$N_{\rm  dot}  =  N/L^2$  for  our  samples  is  about  $10^{10}  {\rm
cm^{-2}}$, and we can simulate systems of $N = 2500 - 10000$ dots. The
positions of the  dots are random,  with one restriction:  no two dots
can be closer to each other than the minimum dot separation $r_{\rm min}$.
The  results  weakly  depend  on  $r_{\rm  min}$;  we  can  deduce its
approximate value $r_{\rm min} = 250 {\rm \AA}$ from the TEM data  for
our samples. Also  at this step  we generate Gaussian  random energies
$\varphi_i$.  \\  2)  Monte-Carlo  step.  On  this  step  we  use  the
Metropolis-type Monte-Carlo algorithm to calculate an average  density
of electrons on the  dots $n = \langle  n_i \rangle$ for a  given gate
voltage $V_g$ and temperature $T$.  
In this paper we have not studied the temperature dependence of the
capacitance, and the 
temperature was maintained at $T = 4.2 \rm \ K$.
We employ the parallel version  of
Metropolis  algorithm  for  Coulomb  systems\cite{Parallel}.  The gate
voltage is being  slowly varied across  the desired interval,  so that
the system  has enough  time to  reach equilibrium  for each  value of
$V_g$.  \\  3)  Calculations  of  capacitance.  In  order  to find the
capacitance $C_{\rm dot}$ of the  dots as a function of  gate voltage,
we compute the derivative
$ \frac{e d n}{d V}$. It is easy to show that the correction to the capacitance 
per unit area can be calculated from the equation

\begin{equation}
C_{\rm dot}(V) = \frac{e d n}{d V}\frac{d}{t}(1-\frac{d}{t}).
\end{equation}
\noindent  The voltage $V$ is connected to the gate voltage $V_g$ by Eq. (1).
This  capacitance  has  a  lot  of  ``noise''  unless   the
temperature is very high  or the number of  dots $N$ is large,  and we
get a smooth curve by convoluting it with a Gaussian filter with small
width. The results of these computations are then compared with the experiment.

The computations were  done on the  Intel Paragon using  128 nodes. It
takes about 6 hours to compute one C-V curve.

\section{Results and Analysis}

\begin{figure}[t]
\epsfxsize=3in
\epsffile{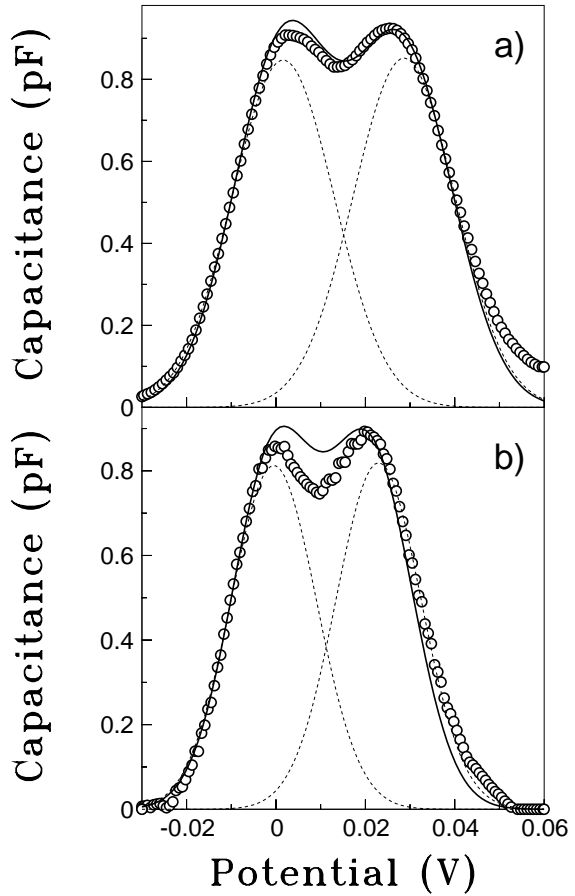}
\caption{Dot capacitance {\it vs\/} potential in the dot plane for two
samples with different distances $d$ between dot and gate planes:  a)
- {\sf A},  $d = 450  {\rm \AA}$, b)  - {\sf B},  $d = 200  {\rm
\AA}$. Circles show the experimental data with background  capacitance
subtracted,  solid  lines  give  the  results  of  our  modeling  with
non-Coulomb disorder $\sigma = 8.5 {\rm meV}$ and dot charging  energy
$W = e^2/\kappa C = 23 {\rm meV}$.
\label{FigComp}}
\end{figure}

We begin by comparing the  experimental CV-curves with the results  of
our  modeling  (Fig.~\ref{FigComp}).  First  of  all,  we subtract the
background capacitance from the experimental data as described  above.
Since the areas  of the real  and simulated devices  are different, so
the  two  capacitances  cannot  be  directly  compared.  We  therefore
normalize  the  calculated  capacitance  so  that  the areas under the
curve, which give the total  charge accumulated on the dots  at large
voltages,  are  the  same.  We  have  checked  that  the  area  of the
experimental capacitance peak corresponds to two electrons per dot for
our sample  sizes and  dot density  $N_{\rm dot}  \approx 10^{10} {\rm
cm^{-2}}$, which  is in  good agreement  with the  TEM measurements of
$N_{\rm dot}$. We also convert the gate voltages to the potentials  in
the dot plane using the lever-arm coefficient Eq.~(\ref{lever}).

The  Fig.~(\ref{FigComp})  shows  the  measured (circles) and computed
(lines) dot capacitances  for two samples  with different dot  to gate
distances $d = 450 {\rm \AA}$ and $d = 200 {\rm \AA}$. The  parameters
of the modeling  were chosen to  get a best  fit for the  sample {\sf
A} with $d =  450 {\rm \AA}$ (Fig.~(\ref{FigComp}a):  dot charging
energy $W  = e^2/\kappa  C =  23 {\rm  meV}$ and  non-Coulomb disorder
magnitude  $\sigma_\varphi  =  8.5  {\rm  meV}$.  We then computed the
capacitance  for  the  second  sample,  {\sf  B} with $d = 200 {\rm
\AA}$,   using    the   same    parameters.   One    can   see    from
Fig.~(\ref{FigComp})b   that   the   theoretical   and    experimental
capacitances agree  quite well.  Since both  samples were  grown under
similar conditions and differ only  in the thickness of the  tunneling
layers, it  is reasonable  to assume  that the  sizes of  the dots and
their fluctuations should be the same for both samples. Therefore, the
ability of our model to reproduce  the effect of the change of  $d$ on
the capacitance,  which is  quite substantial,  without any additional
parameters should be considered as an evidence in favor of the model.

We now turn  to discussing the  effects of the  Coulomb interaction on
the dot  capacitance. First  of all,  the Coulomb  interaction of  two
electrons  in  one  dot  is  responsible  for  the  structure  of  the
capacitance peaks. In the  absence of disorder the  capacitance would
exhibit two delta-function peaks separated by the dot charging  energy.
These peaks are located at  those values of gate voltage  $V_{g1}$ and
$V_{g2}$ at which first and second electrons, respectively, enter  the
dot.  The  corresponding  potentials  in  the  dot  plane  are  $V_1 =
V_{g1}/L$  and  $V_2  =  V_{g2}/L$.  The  disorder broadens the peaks,
however,  they  are  still  observable  in  our samples. The charging
energy depends first and foremost on the  dot shape. It has, however,  a correction,
which comes from the interaction of the charges in the dot with  their
electrostatic images  in the  metallic gate.  It is  easy to show that
this correction  reduces the  voltage between  the capacitance  peaks
$\delta V = V_2 - V_1$ by

\begin{equation}
\delta V_{\rm gate} = - \frac{e}{2 \kappa d}.
\label{dv1}
\end{equation}

\noindent This correction leads to the observed dependence of the peak
splitting on the distance $d$ between the dots and the gate.

This picture is further complicated by the interaction of electrons on
different dots. This interaction gives rise to two effects. First,  in
order  to  load  an  electron  into  a  dot,  one  has to overcome the
repulsion of all charges already  on the dots. This effect  shifts the
position of capacitance peaks to higher voltages. The second  peak
is  shifted  much  more  than  the  first one,  since  the  shift   is
proportional to the charge density already on the dots. Therefore, the
dot-dot interaction increases the splitting between capacitance  peaks
$\delta  V$.  This  correction  to  $\delta  V$, which we call $\delta
V_{\rm int}$, is also $d$-dependent: the smaller is $d$ the weaker  is
the screened Coulomb  interaction and the  smaller is the  correction.
One  can  see  that  the  $d$-dependence  of  $\delta V_{\rm int}$ and
$\delta V_{\rm gate}$ are the same -- the peak splitting increases  at
large $d$ due  to both corrections.  A very rough  estimate of $\delta
V_{\rm  int}$  can  be  obtained  by calculating the average potential
created by a random set of singly charged dots (with minimal  distance
between nearest dots $r_{\rm min}$):

\begin{eqnarray}
\delta V_{\rm int} = && N_{\rm dot} \int V(r)\: d^2 r =
\nonumber \\
&&\frac{2\pi e}{\kappa} N_{\rm dot} \left( \sqrt{4 d^2 + r_{\rm
min}^2} - r_{\rm min}\right).
\nonumber
\end{eqnarray}

\begin{figure}[t]
\epsfxsize=3in
\epsffile{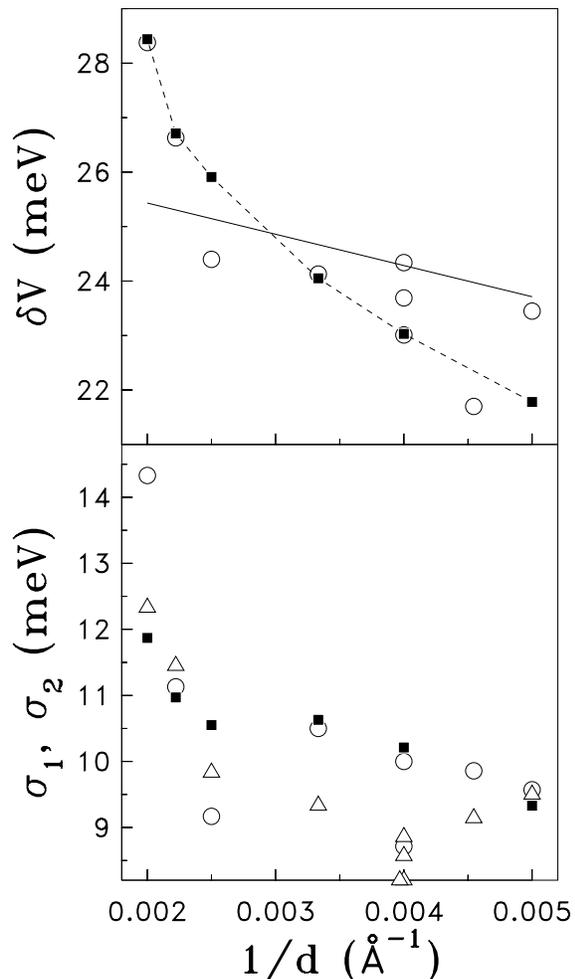}
\caption{Capacitance peaks  splitting $\delta  V$ (a)  and peak widths
$\sigma_1$ and $\sigma_2$ (b) {\it vs\/} the reciprocal dot-gate
distance $1/d$.  The experimental values of  $\delta V$
and  $\sigma_1$  are  shown  by  circles,  those  of  $\sigma_2$  - by
triangles. The results of the modeling are shown by solid squares. The
modeling gives very close values of $\sigma_1$ and $\sigma_2$, so only
$\sigma_1$ are shown. The dashed line is a guide for the eye. The solid
line  shows  the  slope  of  the  correction  $\delta  V_{\rm  gate}$,
Eq.~(\protect\ref{dv1}).
\label{FigDDep}}
\end{figure}

The second effect of the interdot interaction also has to do with  the
potential  created  by  the  charges  on  the dots. Since the dots are
placed randomly in  the plane, the  potential created by  them is also
random. For  each dot,  this potential  is added  to the  ground state
energy, resulting in effectively increasing the total disorder in  the
system. This additional, ``Coulomb disorder'', is also  $d$-dependent.
Its order-of-magnitude estimate is  provided by the dispersion  of the
screened Coulomb potential of randomly placed dots:

\begin{eqnarray}
\delta \sigma_{\rm int}^2 = && N_{\rm dot} \int V(r)^2
\: d^2 r =
\nonumber \\
&&\frac{2\pi e^2}{\kappa^2} N_{\rm dot}\:  {\rm ln} \left\{
\frac{\left[r_{\rm min} + \sqrt{4 d^2 + r_{\rm min}^2}\right]^2}
{4 r_{\rm min} \sqrt{4 d^2 + r_{\rm min}^2}} \right\}.
\nonumber
\end{eqnarray}

\noindent This ``Coulomb disorder'' will add to the disorder caused  by
fluctuations of the size quantization energies in the dots,  resulting
in wider capacitance peaks.

One can easily see the above interaction effects in the modeling.  For
example, the theoretical curve in Fig.~\ref{FigComp}a is obtained  for
charging energy $W  = 23 {\rm  meV}$ and disorder  magnitude $\sigma =
8.5 {\rm  meV}$, while  the distance  between the  peaks is  $\delta V
\approx 24.5 {\rm meV}$ and peak widths are $\sigma_1 \approx \sigma_2
\approx 9.5 {\rm meV}$. However, experimentally all of the effects  of
the Coulomb interaction  are best studied  using their sensitivity  to
the  screening  by  the  metallic  gate.  All  of  these  effects  are
qualitatively evident already in the Fig.~\ref{FigComp}, where one can
see that for the sample {\sf A} with larger $d = 450 {\rm \AA}$ the
peaks are wider and further apart than for the sample {\sf B}  with
smaller $d = 200 {\rm \AA}$.

For  a  quantitative  analysis,  in  Fig.~\ref{FigDDep}  we  compare  the
measured and computed values of  peak splitting $\delta V$ and  widths
$\sigma_1$, $\sigma_2$ for  a set of  samples with different  dot-gate
distances  $d$.  Again,  the  dot  capacitance $C$ and the non-Coulomb
disorder $\sigma$ were chosen from best  fit for one sample with $d  =
450 {\rm \AA}$, and then the results were computed  for other values of  $d$ without
any additional  adjustable parameters.  One can  clearly see  that the
peak splitting increases  with $d$, and  that this increase  is faster
than the  correction $\delta  V_{\rm gate}$  (Eq.~(\ref{dv1})) alone would
give.  This  shows  that  the  dot-dot  Coulomb  interaction, which is
responsible for the  second correction $\delta  V_{\rm int}$, is  very
important in the arrays  of self-assembled quantum dots.  The modeling
includes both of those corrections, and it provides very good
description of the experimental data. The peak widths also increase
with $d$ as a result of the Coulomb interaction (see Fig.~\ref{FigDDep}b),
again with modeling providing reasonably good description of the
experiment.

\section{Conclusions}

We have presented a capacitance spectroscopy study of
InAs self-assembled quantum dots grown on GaAs and compared the
experimental results with the computer simulations. By doing such a 
comparison for a series of devices with different distances between the
dot and gate planes,
we have shown the effect of the Coulomb interactions, both
inside one dot and between the dots, on the capacitance spectra.

We have found out that the Coulomb interaction provides two capacitance peaks which
correspond to sequential loading of two electrons in one dot. The distance 
between the peaks and the width of the peaks  contain an information about 
interactions inside one dot and between the dots.
 In our experiments the interdot interaction was weaker than the external
(non-Coulomb) disorder. A new, interesting physics could be observed in an ordered
 dot array with weak disorder: then the interaction controls the
distribution of electrons on the dots. This may change the sign of the derivative 
 $ \frac{e d n}{d V}$ as was predicted by Efros\cite{17} and was observed by Eisenstein, 
Pfeffer, and West\cite{18} in clean $2D$ systems.
    
\section{Acknowledgments}

We are grateful to W. Kohn for helpful discussions.
This work was supported by the Center for Quantized
Electronic Structures (QUEST) of UCSB, Grant DMR 91-2007, 
and by San  Diego Supercomputer
Center.  F.  G.  P. acknowledges  support    by the NSF Grant DMR 93-08011
 and by the
Quantum Institute of UCSB.
 A. L. E. acknowledges  support   by the Subcontract KK3017 of QUEST of
UCSB.
G. M. R. would like to acknowledge the financial
support
     from CNPq, Brazilian agency.


\begin{references}
\bibitem{WhyOne}
The capacitance $C_i$, in fact, depends on the number of
electrons in the dot, and approaches the conventional geometrical
capacitance when this number is large enough. However, in this paper 
we only consider $n_i \le 2$, so only the capacitance for adding the
second electron is relevant.
\bibitem{whyC} The reason for  neglecting the fluctuations of  the dot
capacitance  is  as  follows:  the  fluctuations  of  $C_i$ would have
broadened  the  capacitance  peak  for  the second electron, while the
first peak would remain unchanged since  one electron in a dot has  no
Coulomb  energy.  The  experiment,  however,  shows that the widths of
first  and  second  peaks  are  very  close.  This  implies  that  the
fluctuations of dot capacitance are negligible.
\end{references}
\end{document}